\documentclass[aps,prl,twocolumn,groupedaddress,longbibliography]{revtex4-1}

\usepackage{graphicx}
\usepackage{dcolumn}

\begin{document}
	
	
	\title{Confinement of high- and low-field-seeking Rydberg atoms using time-varying inhomogeneous electric fields}
	
	
	\author{A. Deller}
    \author{S. D. Hogan}
	\affiliation{Department of Physics and Astronomy, University College London, Gower Street, London WC1E 6BT, United Kingdom}
	
	
	\date{\today}
	
	\begin{abstract}
		Helium atoms in high- and low-field-seeking Rydberg states with linear and quadratic Stark shifts have been confined in two dimensions and guided over a distance of 150~mm using time-varying inhomogeneous electric fields. This was achieved with an electrode structure composed of four parallel cylindrical rods to which voltages were applied to form oscillating and rotating saddle-point fields. These two modes of operation result in time-averaged pseudopotentials that confine samples in high- and low-field-seeking states about the axis of the device. The experimental data have been compared to the results of numerical particle trajectory calculations that include effects of blackbody radiation and electric field ionization. The results highlight important contributions from single-photon blackbody-induced transitions that cause large changes in the principal quantum number of the Rydberg atoms.
	\end{abstract}
	
	\pacs{}
	
	\maketitle
	

The large static electric dipole polarizabilities, and static electric dipole moments of Rydberg states of atoms and molecules allow forces to be exerted on them using inhomogeneous electric fields~\cite{wing80a,breeden81a}. This has led to the development of the methods of Rydberg-Stark deceleration for controlling the motion of matter and antimatter~\cite{hogan16a,cassidy18a}, and experimental tools including deflectors~\cite{townsend01a,yamakita04a,allmendinger14a}, guides~\cite{lancuba13a,ko14a,deller16a}, velocity selectors~\cite{alonso17a}, lenses~\cite{vliegen06a}, mirrors~\cite{vliegen06b,jones17a}, beam splitters~\cite{palmer17a}, decelerators~\cite{yamakita04a,vliegen04a,vliegen05a}, and traps~\cite{vliegen07a,hogan08a,hogan09a,seiler11a,hogan12a,lancuba16a}.

Rydberg-Stark deceleration can be used to prepare cold ($E_{\mathrm{kin}}/k_{\mathrm{B}}\sim100$~mK), quantum-state-selected samples of a wide range of neutral atoms and molecules~\cite{bell09a,hogan11a,meerakker12a}. It has permitted measurements of excited-state decay processes and effects of blackbody radiation on previously inaccessible time scales~\cite{seiler11a,seiler16a}; contributed to new methods for studying ion-molecule reactions at low temperatures~\cite{allmendinger16a,allmendinger16b}; and played a key role in recent advances in positronium (Ps) physics~\cite{deller16a,jones17a,alonso17a}. Rydberg-Stark deceleration also offers further opportunities for studies of slow decay processes of molecular Rydberg states, investigations of resonant-energy and charge transfer in collisions of Rydberg atoms and molecules with polar ground-state molecules~\cite{zhelyazkova17a, zhelyazkova17b} and surfaces~\cite{gibbard15a}, spectroscopic tests of fundamental physics (see, e.g., Ref.~\cite{cassidy18a}), and applications in hybrid quantum information processing~\cite{hyafil04a,rabl06a,hogan12b}. 

Inhomogeneous electric fields have been used to exert forces on samples in high-field-seeking (HFS) Rydberg states, with static electric dipole moments oriented antiparallel to the local electric field and negative Stark shifts, and in low-field-seeking (LFS) states, with dipole moments oriented parallel to the field and positive Stark shifts~\cite{townsend01a, yamakita04a, vliegen05a}. However, with one exception~\cite{ko14a}, all experiments in which Rydberg atoms and molecules were confined in two or three dimensions~\cite{vliegen07a,hogan08a,seiler11a,hogan12a,lancuba13a,deller16a} were performed with LFS states~\cite{hogan16a}. This is because static trapping potentials, with large phase-space acceptances, can be realized in free space about electric field minima for LFS states, but not for HFS states~\cite{ernshaw39a}. However, for several applications it is desirable to confine samples in HFS states. These include (1) studies of stereodynamics in merged-beam collisions of atoms or molecules in low-$\ell$ Rydberg states ($\ell$ is the orbital angular momentum quantum number) with polar ground-state molecules~\cite{zhelyazkova17a,zhelyazkova17b}, (2) experiments with Rydberg Ps and antihydrogen, in which the selective preparation of LFS states is challenging, and simultaneous confinement of HFS and LFS states in guides or velocity selectors is advantageous~\cite{cassidy18a,bertsche14a}, and (3) hybrid quantum information processing with atoms in HFS circular Rydberg states~\cite{hyafil04a,rabl06a,hogan12b}.

The only experiments we are aware of, in which confinement of atoms in HFS Rydberg states was achieved (in two dimensions) were performed using states with linear Stark shifts which resulted in stable orbits in the static $1/r$ electric field surrounding a charged wire~\cite{ko14a}. The phase-space distributions of beams guided in such fields are not optimal for low-temperature merged-beam collision studies, and such confinement can only be achieved for HFS states with linear Stark shifts. An alternative approach to the confinement of samples in HFS states with both linear and quadratic Stark shifts, which can also be used to confine samples in LFS states, is to generate periodic, time-varying electric field distributions which, when applied within an appropriate frequency range, give rise to effective potential minima. Here, we demonstrate, for the first time, confinement of Rydberg atoms in two dimensions using such fields. The methods employed extend from work on guiding and trapping polar ground state molecules~\cite{junglen04a,veldhoven05a,bethlem06a} with adaptations to minimize nonadiabatic transitions between Rydberg states. Two time-varying electric field distributions have been investigated: The first oscillates between two perpendicular saddle-point configurations; the second forms a continuously rotating saddle point. The corresponding pseudopotentials are analogous to those in oscillating~\cite{paul53a,paul90a} and rotating~\cite{hasegawa05a} radio-frequency ion traps.

\begin{figure}
\includegraphics[width=0.45\textwidth]{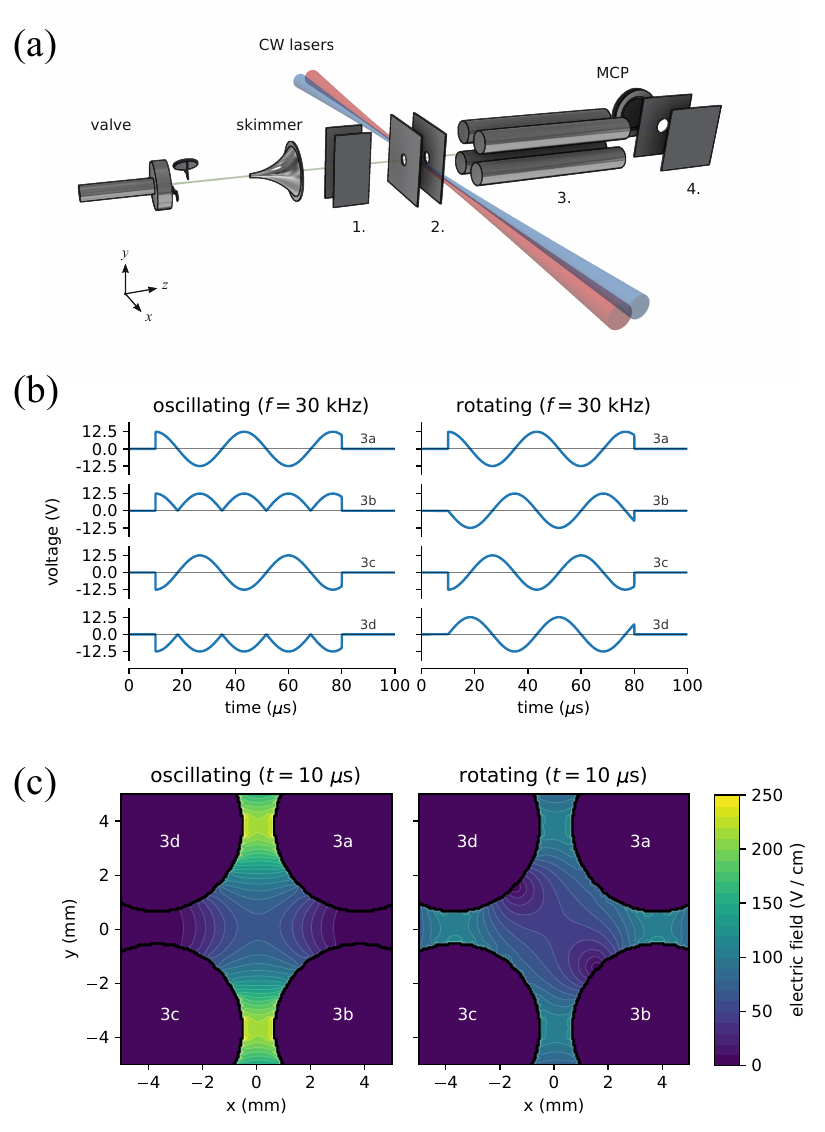}
\caption{ (a) Schematic diagram of the experimental apparatus. Region 1: ion filter; region 2: Rydberg state photoexcitation; region 3: parallel cylindrical electrodes; region 4: detection. (b) Time-dependent voltages for $f=30$~kHz, and (c) electric field distributions at $t=10~\mu$s in the oscillating (left) and rotating (right) modes for $V_{\mathrm{pp}}=25$~V. The contour lines in (c) are indicated in intervals of 10~V\,cm$^{-1}$ with fields on the axis of the electrodes of 67~V\,cm$^{-1}$ (left) and 41~V\,cm$^{-1}$ (right).}\label{fig:expt}
\end{figure}

A schematic diagram of the experiment is presented in Fig.~\ref{fig:expt}(a). A pulsed supersonic beam of He in the metastable 1s2s\,$^3$S$_1$ level ($\overline{v}_z\simeq2000$~m\,s$^{-1}$) was generated in a discharge at the exit of a pulsed valve~\cite{halfmann00a}. After filtering ions formed in the discharge [region 1. in Fig.~\ref{fig:expt}(a)], the atomic beam crossed copropagating cw laser beams used to drive $1\mathrm{s}2\mathrm{s}\,^3\mathrm{S}_1\rightarrow 1\mathrm{s}3\mathrm{p}\,^3\mathrm{P}_2\rightarrow\,1\mathrm{s}37\mathrm{s}\,^3\mathrm{S}/1\mathrm{s}37\mathrm{d}\,^3\mathrm{D}$ transitions~\cite{hogan18a} in the electric field between two parallel copper plates with 3-mm-diameter apertures on the atomic-beam axis [region 2. in Fig.~\ref{fig:expt}(a)]. The lasers were tuned into resonance with the atomic transitions at time $t=0$ for $5~\mu$s [Fig.~\ref{fig:expt}(b)], resulting in the preparation of $\sim10$-mm-long bunches of Rydberg atoms. The atoms then traveled along the axis of four parallel cylindrical metal rods (diameter 6.35~mm, center-to-center spacing 7.45~mm, length 150~mm) [region 3. in Fig.~\ref{fig:expt}(a)]. At $t=10~\mu$s, time-varying voltages were applied to these electrodes for $70~\mu$s. Finally, the Rydberg atoms were detected by ionization in a pulsed electric field of $\sim100$~V\,cm$^{-1}$ [region 4. in Fig.~\ref{fig:expt}(a)]. The resulting He$^+$ ions were collected on a microchannel plate (MCP). The flight time of the atoms from photoexcitation to detection was 100~$\mu$s.


The electric field distributions generated between the four parallel electrodes at $t=10~\mu$s [Fig.~\ref{fig:expt}(b)], for voltages with a peak-to-peak (pp) amplitude $V_{\mathrm{pp}}=25$~V, are presented in Fig.~\ref{fig:expt}(c). In both the oscillating and rotating modes of operation, the field near the axis of the electrodes exhibits a saddle point distribution in the $xy$ plane. In the oscillating mode, atoms in HFS states are initially defocused into the higher field regions above and below $y=0$, whereas the lower field regions on either side of $x=0$ cause focusing in the $x$ dimension. Conversely, atoms in LFS states are focused (defocused) in the $y$ ($x$) dimension. The voltages in Fig.~\ref{fig:expt}(b), applied to electrodes 3a and 3c (3b and 3d), follow sinusoidal (rectified sinusoidal) functions with frequency, $f$. When $t = (2f)^{-1} + 10~\mu$s, the saddle point in Fig.~\ref{fig:expt}(c) is inverted and the focusing and defocusing dimensions are exchanged. In the rotating mode, voltages with a sinusoidal time dependence and relative phase shifts of $90^{\circ}$ are employed [see Fig.~\ref{fig:expt}(b)]. In this case the saddle point continuously rotates and the focusing and defocusfocuing forces are averaged over all angles of orientation of the fields. For any given dipole-moment--to--mass ratio, appropriate combinations of $V_{\mathrm{pp}}$ and $f$ result in two-dimensional confinement of atoms in HFS and LFS states. In the oscillating mode, nonadiabatic transitions between states that are degenerate in zero field can occur. In the rotating mode, the field strength is approximately constant on the axis of the electrodes. Consequently, nonadiabatic transitions can be avoided and samples in states that are degenerate in zero field can be confined.

Along with the experiments, numerical calculations of particle trajectories through the apparatus in Fig.~\ref{fig:expt} were performed. In these calculations, ensembles of Rydberg atoms were generated at the laser photoexcitation position with Gaussian transverse spatial distributions ($\sigma_x=0.3$~mm and $\sigma_y=0.1$~mm) and a uniform distribution (length 10~mm) in the $z$ dimension. The mean velocity of each ensemble was $\vec{v} = (0,0,2000)$~m\,s$^{-1}$, with kinetic energy spreads of $\sigma_{\mathrm{kin}}/k_{\mathrm{B}}=(0.6,7,650)$~mK. Effects of blackbody-induced transitions were included in the calculations using Monte Carlo methods~\cite{seiler11a}. The $n$- and $\Delta n$-dependent blackbody-induced transition rates, $\Gamma(n,\Delta n)$, at 300~K were approximated by exponential functions $\Gamma(n,\Delta n) = (37/n)\,C_1 \exp(-C_2 |\Delta n|^{0.2})$ fit to rates calculated for the corresponding hydrogenlike Rydberg-Stark states~\cite{seiler16a} ($C_1= 2.00\times10^{5}$~s$^{-1}$, $C_2=4.60$ for $\Delta n <0$; and $C_1= 3.09\times10^{5}$~s$^{-1}$, $C_2=4.95$ for $\Delta n >0$). Since $n$-changing transitions between Rydberg-Stark states are strongest between sublevels with similar static electric dipole moments, it was assumed that the dipole moments of the atoms did not change when these transitions occurred. At each calculation time step the electric field experienced by each atom was monitored. If this exceeded $F_{\mathrm{ion}}^{\mathrm{HFS}}(n)= F_{\mathrm{class}}(n)=5.14\times10^{9}/(9n^4)$~V\,cm$^{-1}$ [$F_{\mathrm{ion}}^{\mathrm{LFS}}(n)=2\,F_{\mathrm{class}}(n)$] for the outer HFS [LFS] Stark states~\cite{damburg79a,gallagher94a}, the atom was removed. 


\begin{figure}
\includegraphics[width=0.45\textwidth]{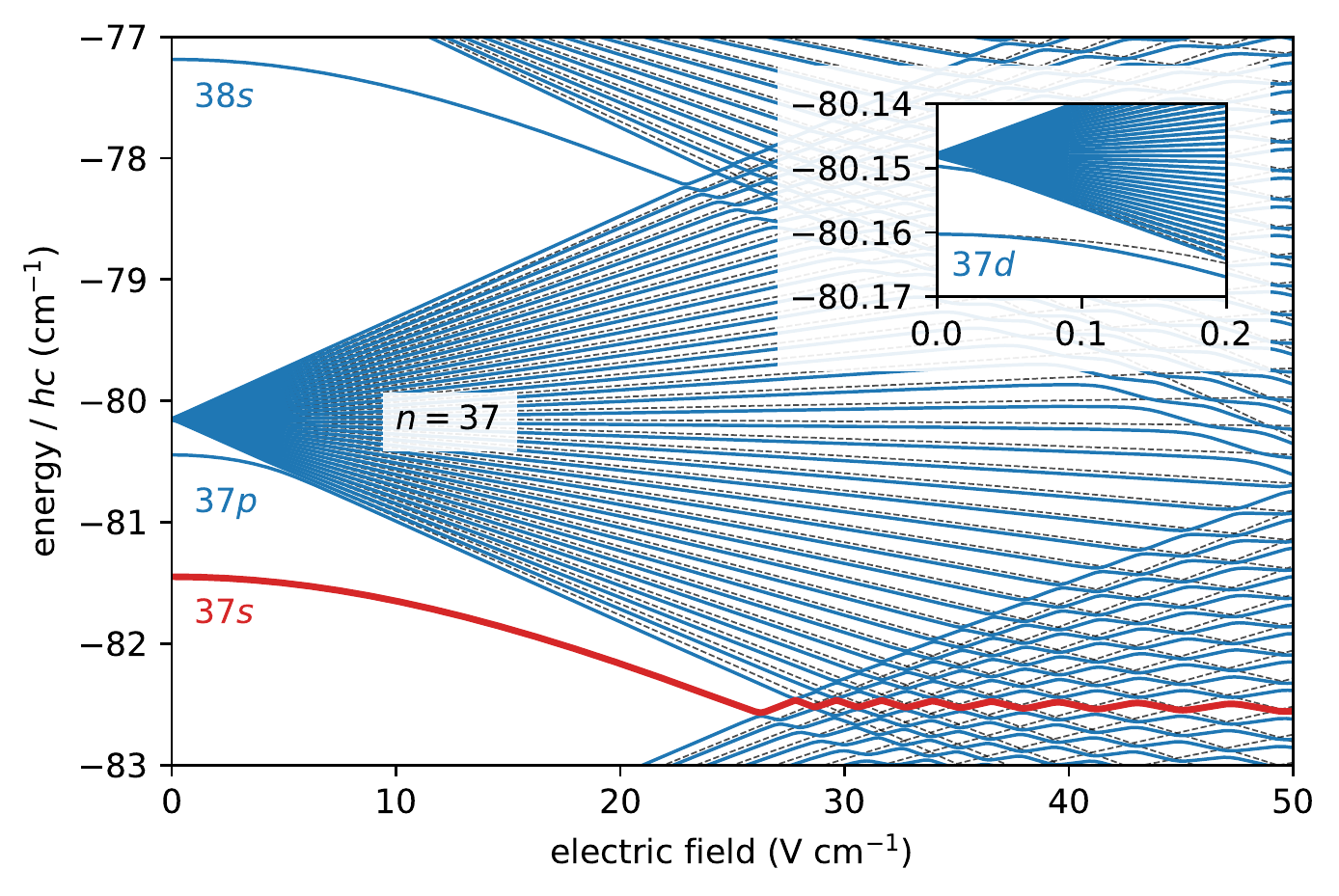}
\caption{Stark map of triplet Rydberg states in He with values of $n$ close to 37. Sublevels with $m=0$ ($|m|=2$) are indicated by the continuous (dashed) curves. The adiabatic path of the 37s state is indicated by the thick red curve. Inset: An expanded view of the region encompassing states with $n=37$ and $\ell\geq2$ in fields below 0.2~V\,cm$^{-1}$.}\label{fig:Stark}
\end{figure}
		
To study the confinement of atoms in Rydberg states with a range of characteristics in the inhomogeneous time-varying electric fields, HFS and LFS states with linear and quadratic Stark shifts were prepared. The HFS 37s (quantum defect $\delta_{37\mathrm{s}}=0.296\,685$) and 37d ($\delta_{37\mathrm{d}}=0.002\,887$) states were employed~\cite{drake99a}. As seen in Fig.~\ref{fig:Stark}, these states exhibit quadratic Stark shifts in weak fields and are nondegenerate with other higher-$\ell$ states in zero field. In fields above the Inglis-Teller (IT) limit~\cite{gallagher94a}, $F_{\mathrm{IT}}$, the 37s state with $m=0$ evolves adiabatically through large avoided crossings. On the other hand, the 37d state, prepared with predominantly $|m|=2$ character, evolves diabatically. Experiments were also performed with atoms in the outer hydrogenlike LFS Stark state with $n=37$ and $k=n_1-n_2=+34$ ($n_1$ and $n_2$ are parabolic quantum numbers~\cite{gallagher94a}). This state has a linear Stark shift, even in fields that approach zero, in which it is highly degenerate with other parabolic states.

\begin{figure}
\includegraphics[width=0.45\textwidth]{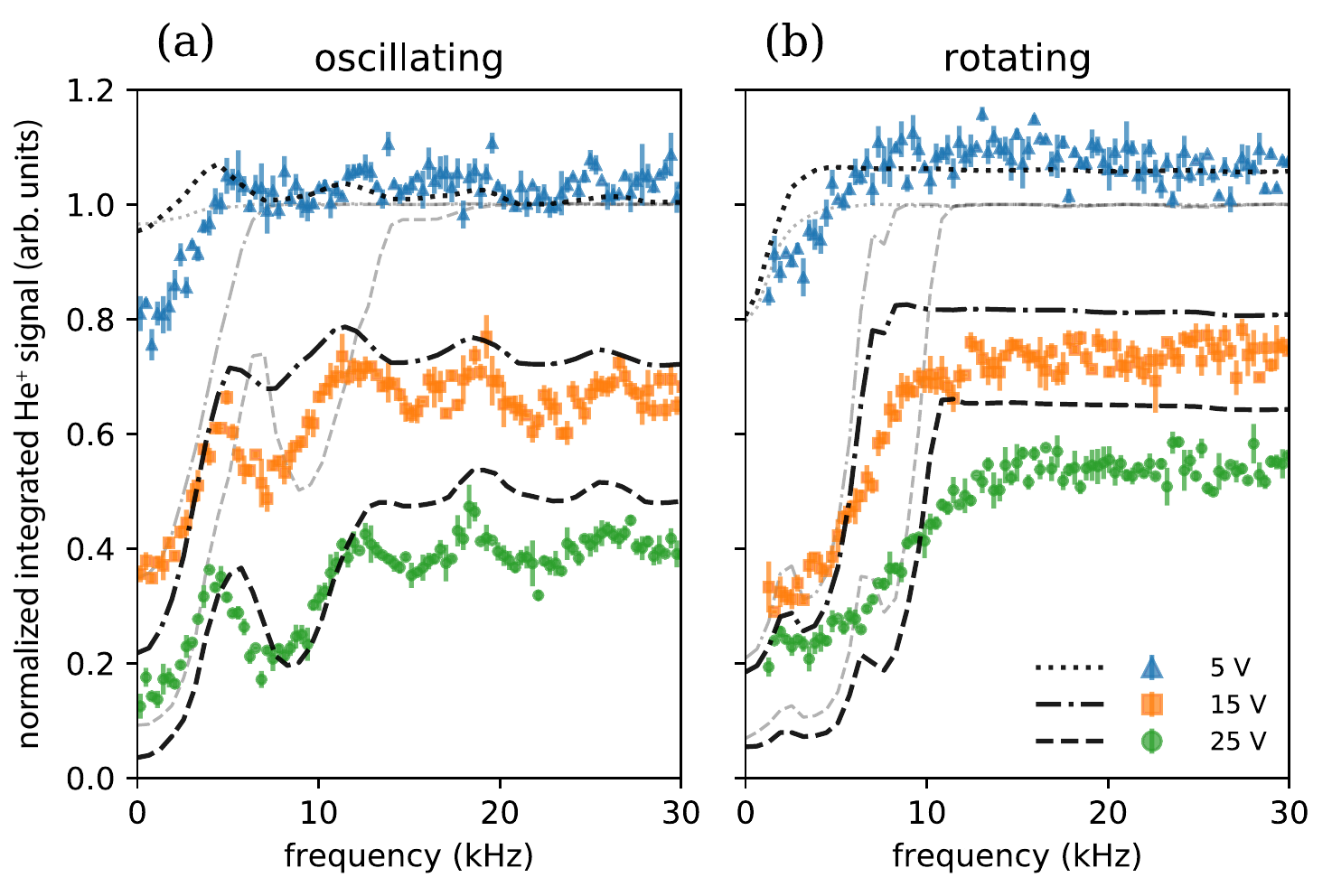}
\caption{Normalized fraction of atoms transmitted to the detection region after photoexcitation of the HFS 37d state for the (a) oscillating, and (b) rotating mode of operation. The measured (points) and calculated (broken curves) data were obtained for $V_{\mathrm{pp}}=5$, 15 and 25~V as indicated in (b). The thin grey (thick black) curves correspond to calculations performed excluding (including) radiative effects.}\label{fig:ddata}
\end{figure}

The measured and calculated fractions of atoms transmitted through region 3. and detected in region 4. of the apparatus, normalized to the signal recorded with the fields switched off, after photoexcitation of the HFS 37d state are displayed in Fig.~\ref{fig:ddata} for the oscillating and rotating modes of operation with $V_{\mathrm{pp}}=5$, 15 and 25~V. The calculations were performed for a field-independent electric dipole moment of the 37d state of $\mu_{37\mathrm{d}}=4\,650$~D. The significance of radiative processes including blackbody transitions is evident from the results of the calculations performed excluding (thin grey curves) and including (thick black curves) these effects. 

In both modes of operation, when $f<5$~kHz less than one oscillation of the voltages occurs during the time in which the fields affect the particle dynamics. Consequently, defocusing of the HFS atoms away from the axis of the electrodes in the $y$ dimension [see Fig.~\ref{fig:expt}(c)] is not compensated as the fields evolved in time. Therefore, the normalized fraction of atoms transmitted to the detection region is low and reduces as $V_{\mathrm{pp}}$ is increased. For $f>5$~kHz, less defocusing occurs in the $y$ dimension and the fraction of atoms detected is significantly larger than for $f<5$~kHz. The increase in the fraction of atoms transmitted to the detection region for $f>10$~kHz and $V_{\mathrm{pp}}=5$~V results from the electric-field dependence of the fluorescence lifetime of the 37d state~\cite{gallagher94a}. This changes from 23~$\mu$s in zero field to 90~$\mu$s in the typical stray fields in the apparatus of $\sim0.25$~V\,cm$^{-1}$, and $300~\mu$s in fields beyond $0.9$~V\,cm$^{-1}$, and was included in the calculations (dashed curves). In general the results of the calculations are in good qualitative agreement with the experimental data.

In the oscillating mode, the greater forces on the atoms when $V_{\mathrm{pp}}=15$ and 25~V give rise to large amplitude transverse motions, and resonances in the detected signal when $f=5$, 12 and 19~kHz. These resonances, seen in the experiments and calculations, are particular to the combination of the initial phase-space properties of the Rydberg atoms and the geometry of the detection region in the experiments, and occur when the time-averaged transverse forces focus the atoms to the detection region. Focusing effects in the rotating mode are averaged over a range of angles, and resonances are not observed. In Fig.~\ref{fig:ddata}(b) the magnitude of the transmitted signal increases as the value of $f$ is increased, with good qualitative agreement between the experimental and calculated data. The threshold between $5$ and $10$~kHz is not as sharp in the experimental data as it is in the results of the calculations because effects of blackbody-induced changes in the dipole moments of the atoms, which accompany $n$ changing, are not included in the calculations.

In Fig.~\ref{fig:ddata}(a) and~(b) an overall reduction in the fraction of detected atoms is observed for $f>10$~kHz as $V_{\mathrm{pp}}$ is increased from 5~V to 25~V. From the calculations, these losses are attributed to effects of electric field ionization of high-$n$ states following blackbody-induced transitions. Detection using a low ionization field (100~V\,cm$^{-1}$) established that at the time of ionization a measurable fraction of the atoms resided in states with values of $n\geq49$. These states are predominantly populated following single-photon blackbody transitions for which $\Delta n\gtrsim10$. Such transitions also transfer population to higher $n$ states that ionize before reaching the detection region. This observation, and the qualitative agreement between the experiments and calculations for $f>10$~kHz, which was only achieved after accounting for these large $\Delta n$ blackbody transitions, highlights the importance of the long tail in the distribution of blackbody transition rates.

\begin{figure}
\includegraphics[width=0.45\textwidth]{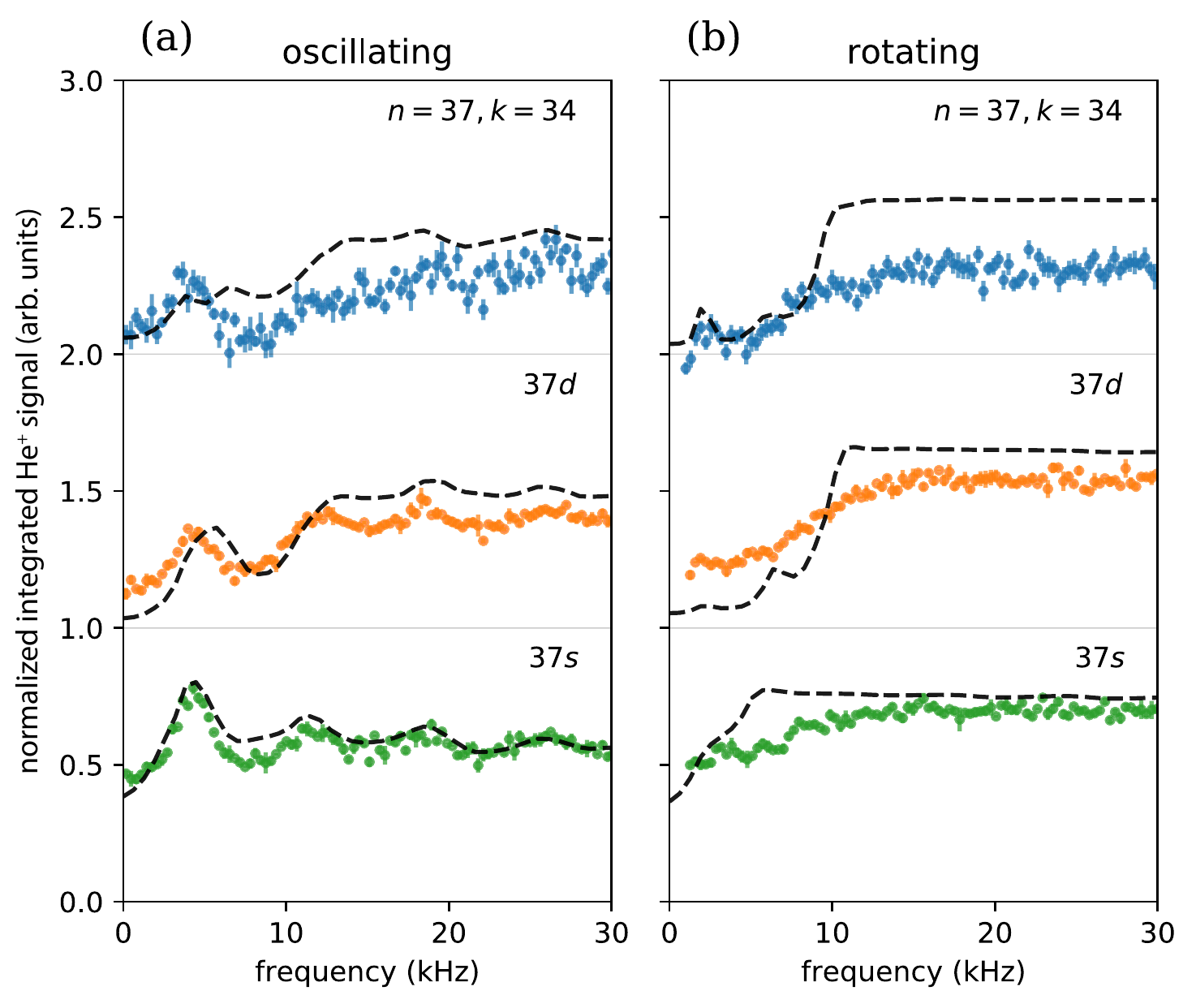}
\caption{Normalized fraction of atoms transmitted to the detection region after photoexcitation of the HFS 37s and 37d states, and the LFS $n=37$, $k=+34$ state for (a) the oscillating, and (b) the rotating mode of operation with $V_{\mathrm{pp}}=25$. The experimental (calculated) data are indicated by the points (broken curves). For clarity of presentation the upper sets of data in each panel are vertically offset.}\label{fig:highfielddata}
\end{figure}

Studies performed with atoms in the HFS 37s state, and LFS $n=37$, $k=+34$ Stark state, for $V_{\mathrm{pp}}=25$~V are presented in Fig.~\ref{fig:highfielddata} and compared to the results obtained for the 37d state under similar conditions. The 37s state has a static electric dipole polarizability of $\alpha_{37\mathrm{s}}=8.3\times10^{-30}$~C\,m$^{2}$/V and exhibits a quadratic Stark shift in fields up to $F_{\mathrm{IT}}\simeq26$~V\,cm$^{-1}$ (see Fig.~\ref{fig:Stark}). In the time-varying fields in the experiments it evolves adiabatically beyond $F_{\mathrm{IT}}$. On average it remains HFS up to the field of $\sim200$~V\,cm$^{-1}$ in which it ionizes. In fields between $F_{\mathrm{IT}}$ and 70~V\,cm$^{-1}$, and from 85 to 112~V\,cm$^{-1}$, $\overline{\mu}_{37\mathrm{s}}=0$~D. In fields from 70 to 85~V\,cm$^{-1}$, and 112 to 200~V\,cm$^{-1}$, $\overline{\mu}_{37\mathrm{s}}=1\,500$~D. These average electric dipole moments were employed in the trajectory calculations for this state. The reduction in the fraction of atoms detected when $f<5$~kHz after preparation of the 37s state -- in both the oscillating and rotating modes -- is not as significant as for the 37d state. This is because the 37d state exhibits a constant static electric dipole moment in all fields in the experiments. The resonance behavior seen in the oscillating mode for the 37d state persists in the studies performed with the 37s state, as does the almost frequency independent transmission of atoms to the detection region when $f>10$~kHz

The LFS $n=37$, $k=+34$ state has an electric dipole moment of 4\,800~D, which is almost equal in magnitude to $\mu_{37\mathrm{d}}$. However, in contrast to the 37d state it is degenerate in zero field. Consequently, it can undergo nonadiabatic transitions and change the sign of its dipole moment in fields that evolve through zero. The experimental data in Fig.~\ref{fig:highfielddata}, recorded after preparation of this LFS state, follow the same general trends as those for the 37d state. The normalized fraction of atoms detected when $f<3$~kHz is similar and, in the oscillating mode, a resonance is observed at $f\simeq6$~kHz. This resonance is less prominent in the calculations. However, because it is observed in the experimental and calculated data for the 37d state, its appearance in the data recorded for the $n=37$, $k=+34$ state indicates nonadiabatic evolution in fields close to zero. Since the on-axis field in the rotating mode of operation is approximately constant over time, nonadiabatic dynamics are not expected.

In conclusion, we have demonstrated two-dimensional confinement of Rydberg atoms using time-varying inhomogeneous electric fields. Experiments were performed with samples in HFS and LFS states with linear and quadratic Stark shifts. Comparison of the experimental data with the results of particle trajectory calculations provided valuable insights into effects of blackbody radiation in the experiments and demonstrated that on long timescales, blackbody-induced transitions for which $\Delta n\gtrsim10$ play an important role. In addition to $m$ mixing~\cite{seiler16a}, such transitions must be considered in the interpretation of these and other Rydberg-Stark deceleration and trapping experiments. The time-varying electric field distributions and electrode structures employed here are ideally suited to merged beam collision experiments with He Rydberg atoms and ground-state NH$_3$~\cite{zhelyazkova17a,zhelyazkova17b}. They are of interest in experiments with Rydberg Ps and antihydrogen as they offer a way to simultaneously confine HFS and LFS states. In a cryogenic environment, where blackbody effects are reduced, an extended version of the device described here is expected to exhibit narrower resonance features and may be used to filter Rydberg atoms or molecules according to their dipole-moment-to-mass ratio for collision studies or to enhance state-selective detection. 

\begin{acknowledgments}
We thank John Dumper for his help constructing the apparatus used here. This work is supported by the European Research Council (ERC) under the European Union's Horizon 2020 research and innovation program (Grant No. 683341).
\end{acknowledgments}

\end{document}